\documentclass[11pt]{article}
\usepackage[margin=1in]{geometry}
\usepackage{graphicx}

\def\beq{\begin{equation}}
\def\eeq#1{\label{#1}\end{equation}}
\def\eeqn{\end{equation}}

\def\beqa{\begin{eqnarray}}
\def\eeqa#1{\label{#1}\end{eqnarray}}
\def\eeqan{\end{eqnarray}}

\let\bar=\overbar

\def\Dslash{\not{\hbox{\kern-4pt $D$}}}
\def\dslash{\not{\hbox{\kern-2pt $\del$}}}

\def\msb{{\bar{\ssstyle M \kern -1pt S}}}

\def\Title#1{\begin{center} {\Large {\bf #1} } \end{center}}
\def\Author#1{\begin{center} {\normalsize {\sc #1} } \end{center}}
\def\Institution#1{\begin{center} {\normalsize {\it #1} } \end{center}}
\def\Abstract#1{\noindent {\normalsize {\bf Abstract:} {\normalfont #1}}}
\def\Conference{\vspace{4mm}\begin{raggedright} {\normalsize {\it Talk presented at the 2019 Meeting of the Division of Particles and Fields of the American Physical Society (DPF2019), July 29--August 2, 2019, Northeastern University, Boston, C1907293.} } \end{raggedright}\vspace{4mm}}

\begin{document}

%
%

\Title{The Heavy Photon Search Experiment}

\Author{Cameron Bravo on Behalf of the HPS Collaboration}

\Institution{SLAC National Accelerator Laboratory, Menlo Park, CA 94025, USA}

\Abstract{The Heavy Photon Search (HPS) experiment searches for an electro-produced dark photon using an 
electron beam provided by the CEBAF accelerator at the Thomas Jefferson National Accelerator Facility. 
HPS has successfully completed two engineering runs. In 2015 using a 1.056 GeV, 50 nA electron beam, 
1.7 days (10 mC) of data was obtained and 5.4 days (92.5 mC) of data was collected in 2016 using a 2.3 GeV, 
200 nA electron beam. In addition, HPS will complete its first physics run in the summer of 2019. HPS looks 
for dark photons through two distinct methods, a resonance search in the $e^{+}e^{-}$ invariant mass 
distribution above the large QED background (large dark photon-SM particles coupling region) and a 
displaced vertex search for long-lived dark photons (small coupling region). HPS employs a compact 
spectrometer, matched to the forward kinematic characteristics of A$^\prime$ electro-production. The detector 
consists of a silicon tracker for momentum analysis and vertexing and a lead tungstate (PbWO$_4$) electromagnetic calorimeter 
for particle ID and triggering. Both analyses are complete for the 2015 engineering run and demonstrate 
the full functionality of the experiment that will probe hitherto unexplored parameter space with 
higher luminosity runs. Results from the 2015 dataset will be presented as well as an update on 
2016 analysis and the status of the 2019 physics run.}

\Conference

%
%

\section{Introduction}

\begin{figure}[]
\centering
\includegraphics[height=2.0in]{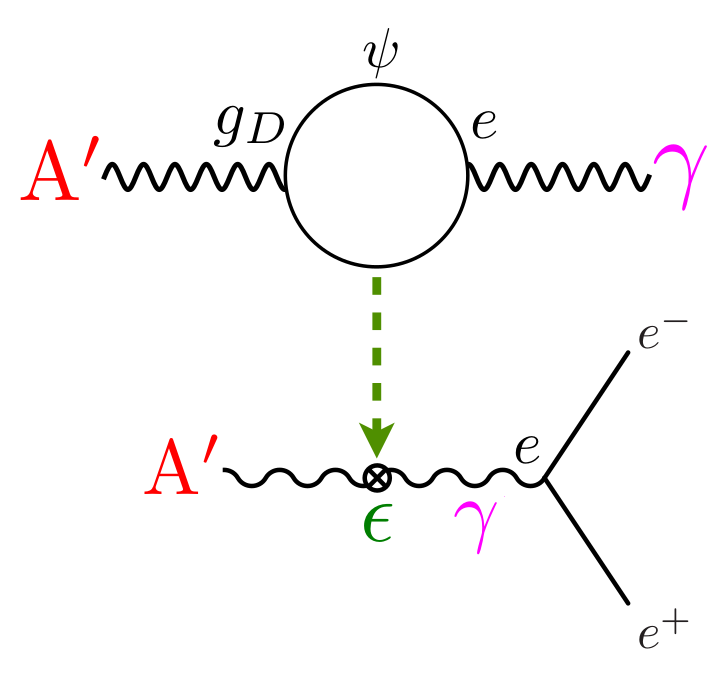}
\caption{The dark sector could include massive fermions which might couple to the SM photon. This would kinematically mix the A$^\prime$ and the SM photon.}
\label{fig:kineMix}
\end{figure}

\begin{figure}[]
\centering
\includegraphics[height=1.5in]{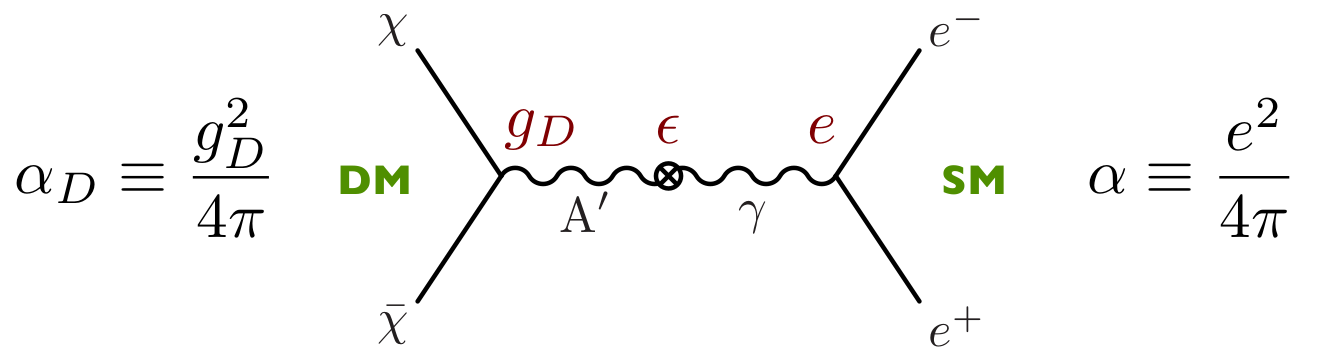}
\caption{The mixing of the photon and A$^\prime$ would couple DM to the SM.}
\label{fig:DMtoSM}
\end{figure}

There has been strong evidence for the existence of Dark Matter (DM) for several decades,
despite the fact that nothing is known about the particle nature of DM.
Weakly interacting massive particles (WIMPs) have been a well motivated candidate, however
the mass of the particles in this scenario have a lower bound \cite{WIMP}.
Lighter thermal relics are also a great candidate but require a new physical force to achieve the
correct relic abundance. We consider the case where DM interacts via a vector mediator \cite{Holdom}.
This new vector mediator could kinematically mix with the photon of the Standard Model (SM),
as shown in figure \ref{fig:kineMix}. This would allow for the interaction of the dark sector with
the SM via electromagnetism suppressed by a mixing parameter $\epsilon$, as shown in figure \ref{fig:DMtoSM}.
This new boson is commonly called a Dark Photon (A$^\prime$).

HPS is a fixed target experiment consisting of a silicon tracker for momentum analysis and 
vertexing and a PbWO$_4$ electromagnetic calorimeter for particle ID and triggering \cite{ECalNIM}.
The experiment employs an electron beam and tungsten target to probe A$^\prime$ scenarios which decay 
exclusively to electron-positron pairs. Two separate analyses are performed to search
for signals which have either prompt or displaced decays.

\section{Dark Photons in Fixed Target Experiments}
\begin{figure}[]
\centering
\includegraphics[height=1.7in]{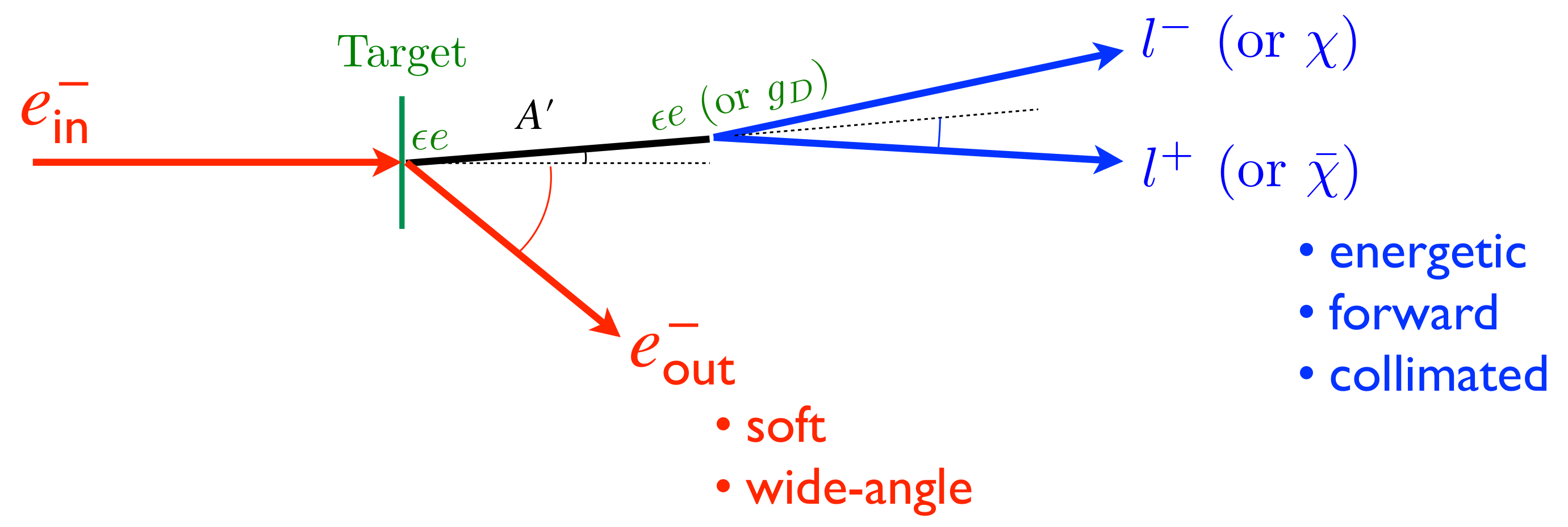}
\caption{A cartoon of A$^\prime$ production in fixed target experiments.}
\label{fig:fixedTargetCartoon}
\end{figure}

\begin{figure}[]
\centering
    \includegraphics[height=1.7in]{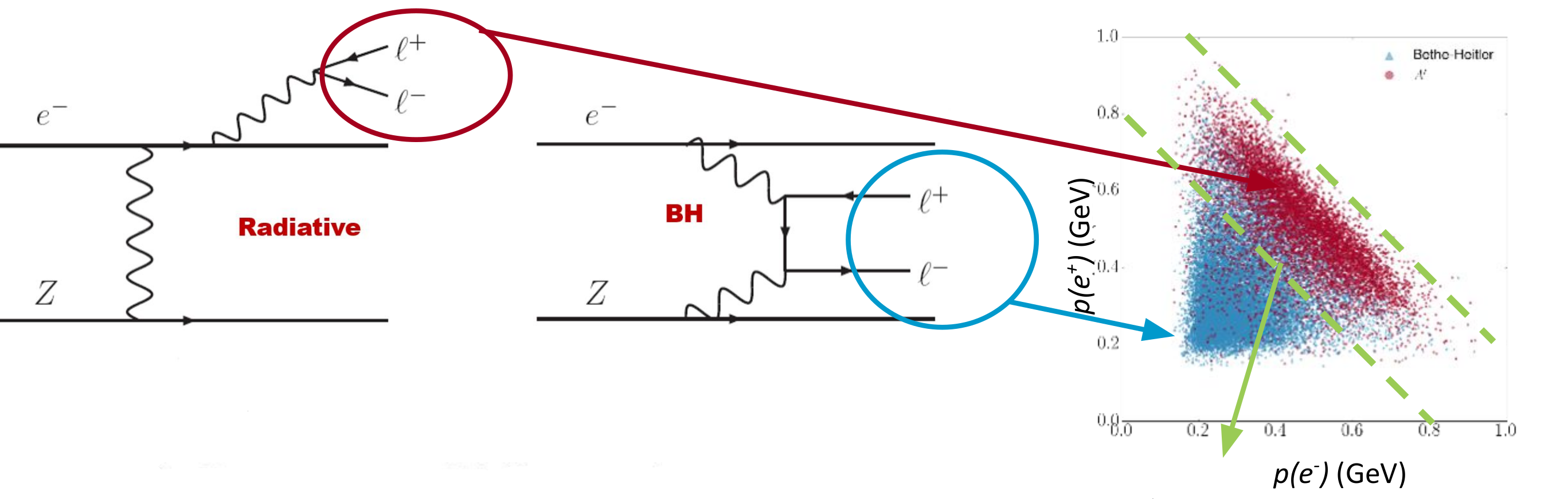}
\caption{The two dominant backgrounds in fixed target experiments searching
    for visible A$^\prime$ decays are radiative tridents and Bethe-Heitler (BH).
    The BH background can be suppressed with a simple kinematic cut on the
    electron and positron momenta.}
\label{fig:hpsBG}
\end{figure}

\begin{figure}[]
\centering
    \includegraphics[height=3.0in]{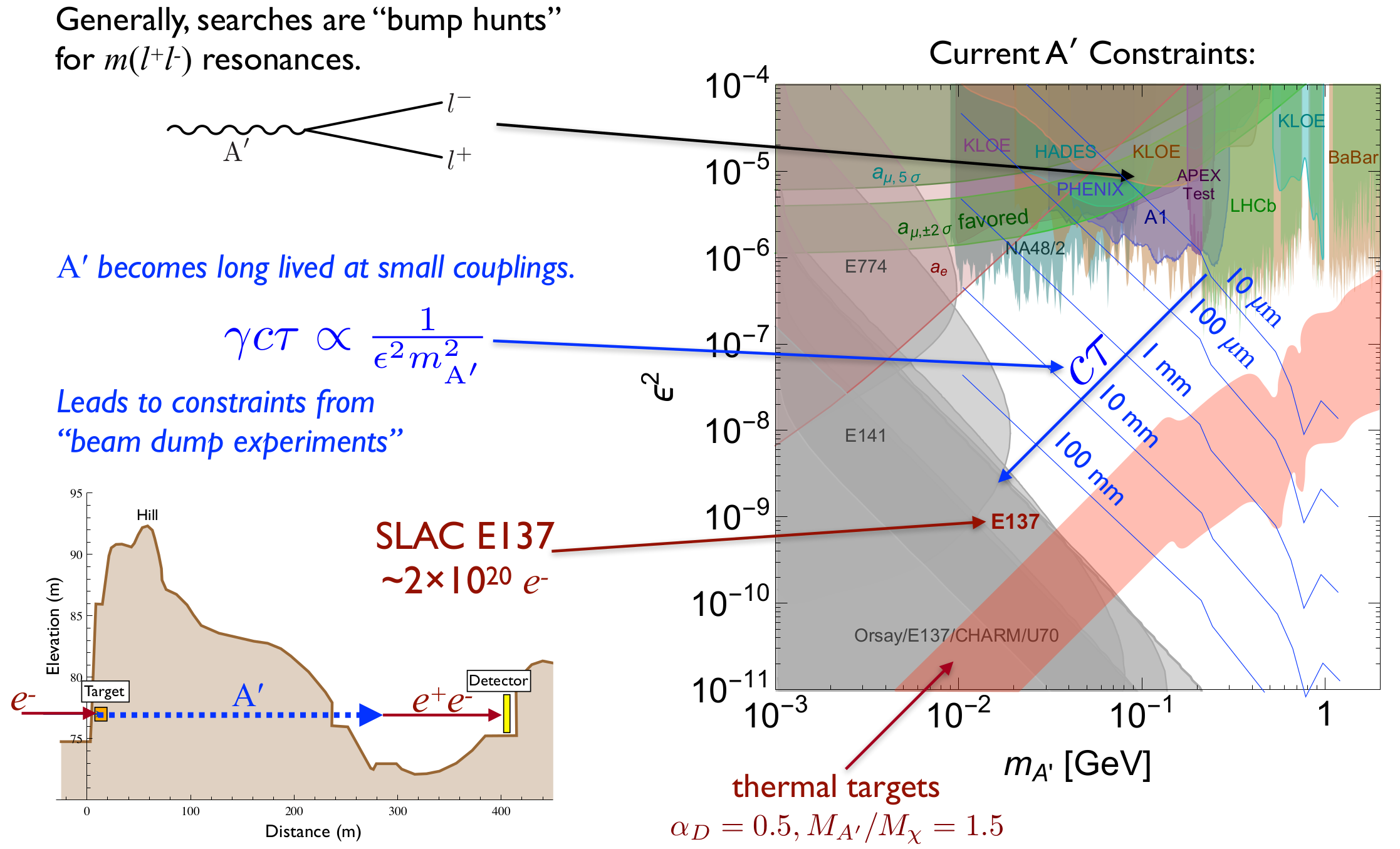}
    \caption{This is the parameter space used to summarize searches for
    visible decays of Dark Photons. $\epsilon^2$ over $10^{-6}$ has
    already been almost entirely excluded. Low-mixing and low-mass scenarios
    have also been excluded by multiple beam dump experiments. HPS is a search
    designed to be sensitive in the most difficult part of the parameter space
    to probe, where the A$^\prime$ lifetime is on the order of mm.}
\label{fig:parSpaceIntro}
\end{figure}

\begin{figure}[]
\centering
    \includegraphics[height=2.3in]{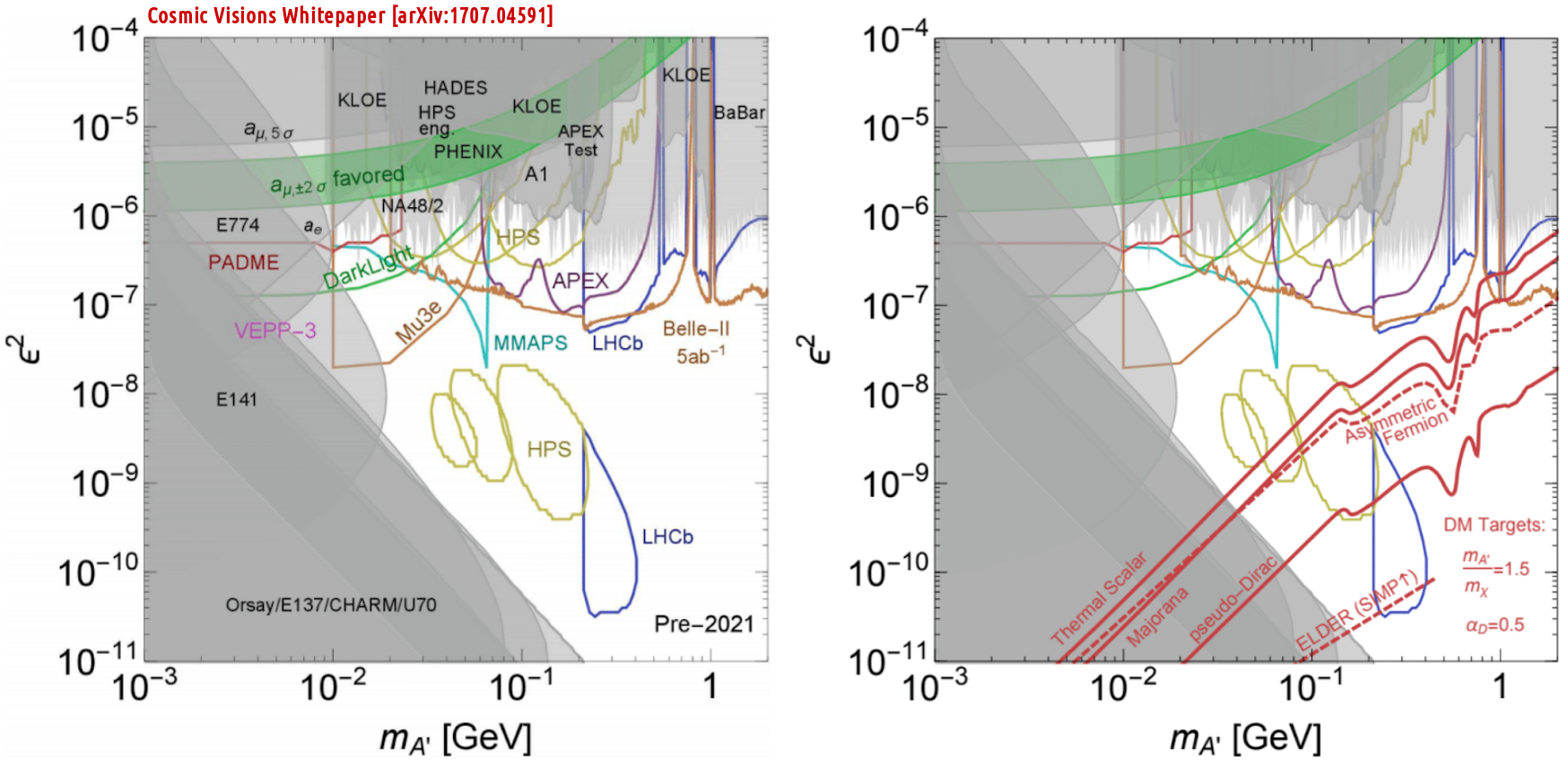}
    \caption{A summary of excluded regions and projections of future experiments
    searching for visible decays of A$^\prime$. This figure was taken from \cite{CosmicVisions}.}
\label{fig:hpsReach}
\end{figure}

Fixed target experiments are well suited to search for sub-GeV Dark Photons
due to very large production cross sections for low-mass states.
\cite{aprime}
The kinematics of A$^\prime$ production result in an A$^\prime$ with most of the beam energy and a soft recoil
electron as shown in figure \ref{fig:fixedTargetCartoon}.
The two dominant backgrounds for visible decays of Dark Photons are
radiative tridents and Bethe-Heitler tridents (BH), shown in figure \ref{fig:hpsBG}. 
The radiative trident rate provides a reference for the expected signal rate.
The signal is typically parameterized by the mass of the A$^\prime$ and
the mixing parameter, $\epsilon$, to the SM photon. The lifetime of the A$^\prime$ has a complicated
relationship to these parameters, which is described in figure \ref{fig:parSpaceIntro}.
The projections of existing limits and upcoming experiments with projected sensitivity 
to this physics are shown in figure \ref{fig:hpsReach}.

\section{The HPS Experiment}
\begin{figure}[]
\centering
    \includegraphics[height=2.8in]{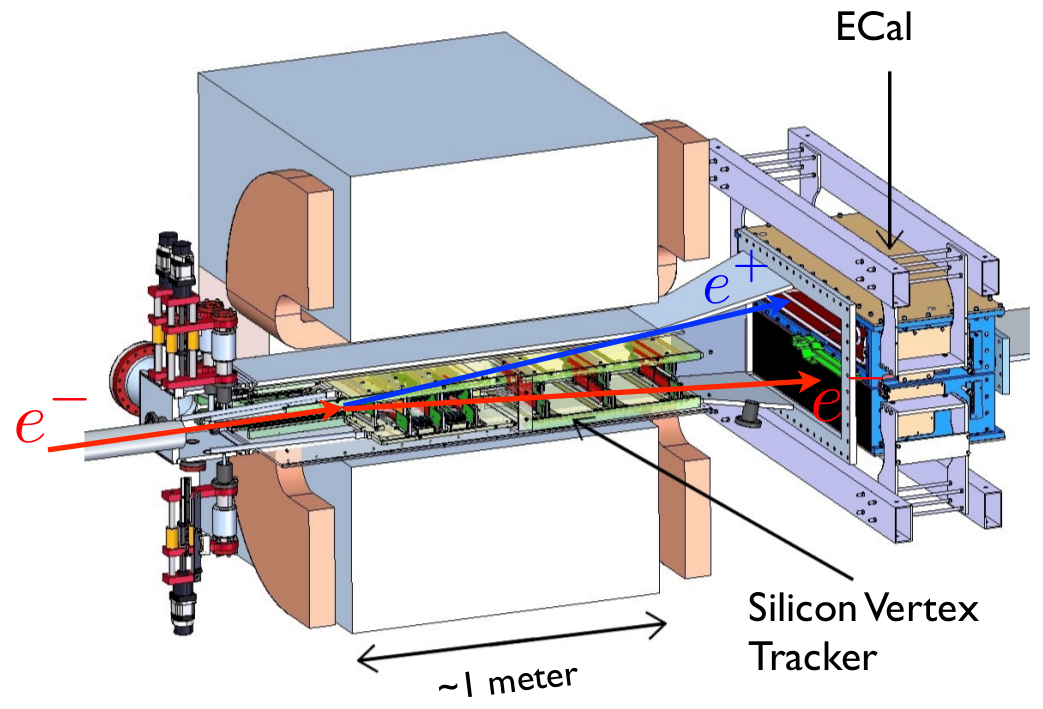}
    \caption{A graphic of the HPS apparatus showing the path of an incoming beam electron and 
    an $e^+e^-$ pair coming from the target.}
\label{fig:hpsDet}
\end{figure}

The HPS apparatus is located in Hall B of the Continuous Electron Beam Accelerator Facility (CEBAF)
at the Thomas Jefferson National Accelerator Facility of the Department of Energy.
The experiment is designed as a general search for visible decays of the A$^\prime$.
The apparatus is a compact $e^+e^-$ spectrometer immediately downstream of a thin target
in a multi-GeV electron beam. It consists of a $PbWO_4$ calorimeter (ECal) and a low-mass,
high-rate (up to 4 MHz/$mm^2$) silicon tracker (SVT). A drawing of the
detector is shown in figure \ref{fig:hpsDet}. The ECal is used as a trigger to eliminate
order 10 MHz of scattered single electrons from the beam. The tracker was
designed with 6 layers of axial-stereo pairs of strip sensors read out by the APV25 ASIC designed for the CMS experiment \cite{apv25}. It is capable of reconstructing vertices several mm downstream of the target. The SVT is supported by kinematic mounts that allow it to be opened and closed for beam tuning operations. The
nominal running position of the tracker leaves a 1 mm gap between the sensors of the first layer.

\subsection{Engineering Runs}
\begin{figure}[]
\centering
    \includegraphics[height=2.5in]{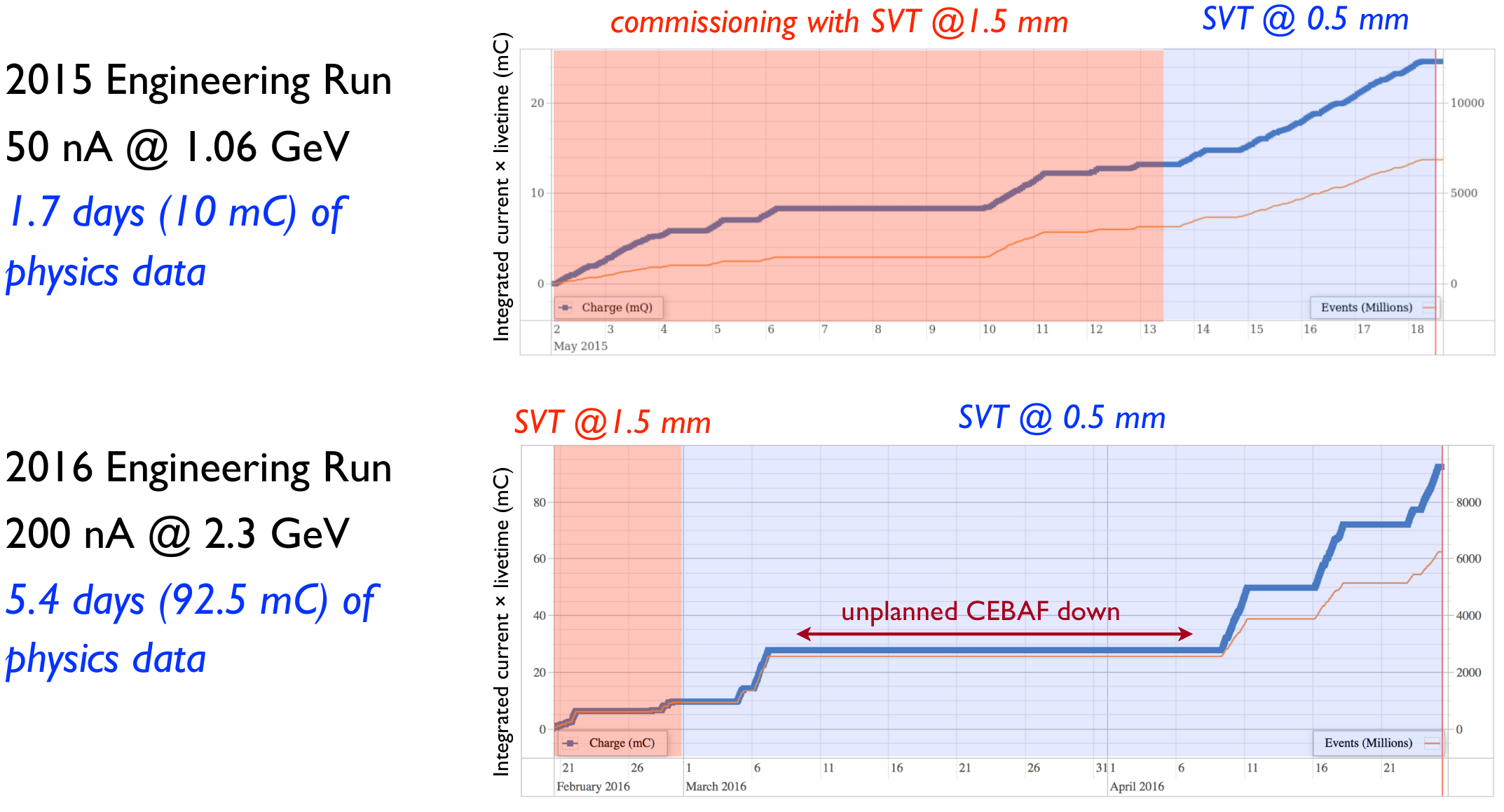}
    \caption{The integrated beam current and total number of triggered 
    events throughout the two HPS engineering runs in 2015 and 2016.}
\label{fig:engRuns}
\end{figure}

HPS has performed two engineering runs, one in 2015 with a beam energy of 1.06 GeV, and one in
2016 with a beam energy of 2.3 GeV. These runs are summarized in figure \ref{fig:engRuns}. The apparatus
has performed exceptionally well, and is still approved for 165 more days of beam time.
The first physics run is underway during this summer (2019) with a beam energy of 4.56 GeV.

\begin{figure}[]
\centering
    \includegraphics[height=2.7in]{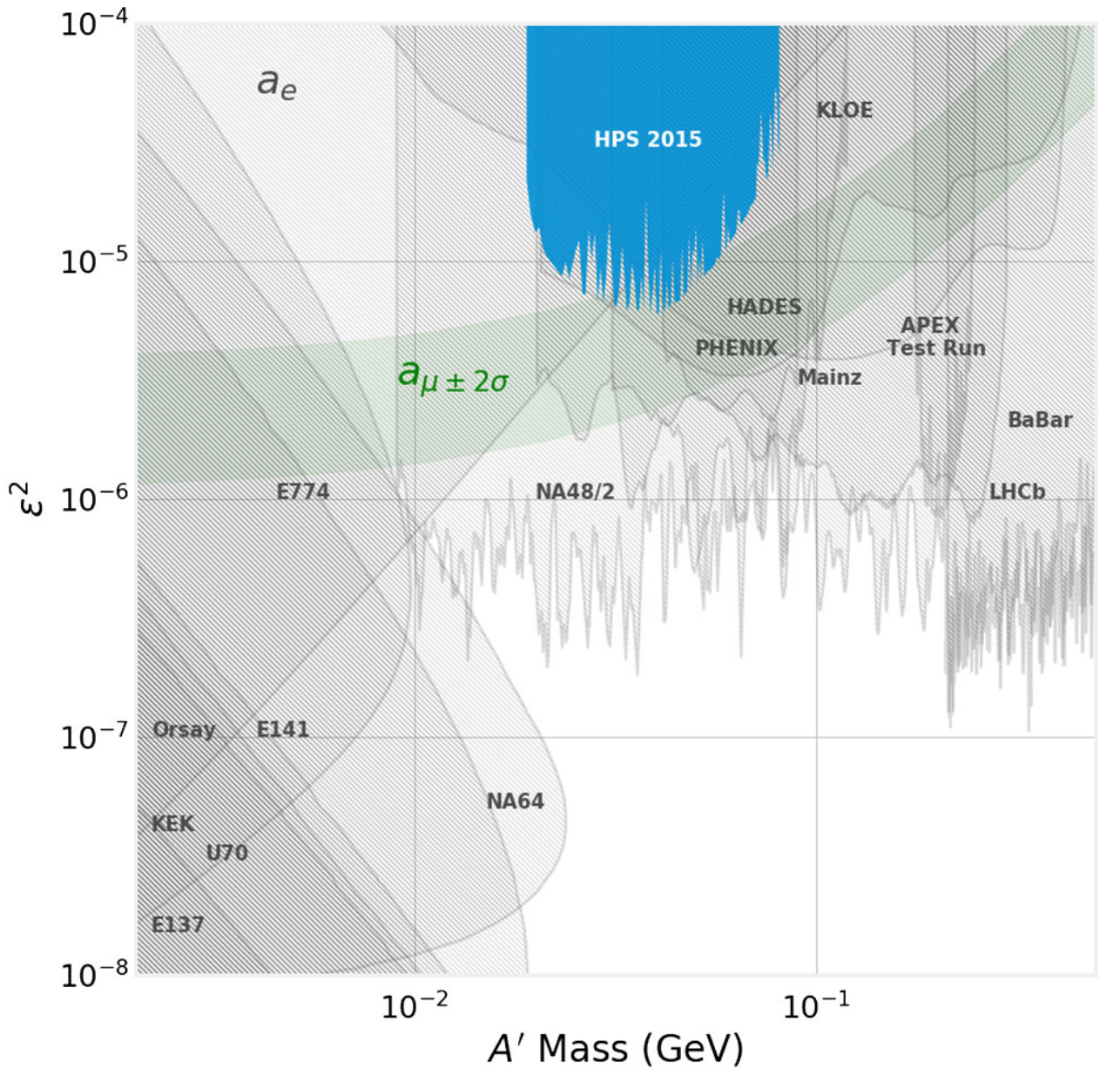}
    \caption{The region of the A' parameter space excluded by the 2015 resonance search analysis.}
\label{fig:hps15limit}
\end{figure}

The results of the resonance search performed on the 2015 dataset have been published \cite{res15}.
This analysis measures the mass resolution using M{\o}llers and the expected signal rate is
calibrated by using tridents. The background shape is fit using a Chebyshev polynomial and
a likelihood ratio is used to quantify the significance of any excess. No signal was observed
so the likelihood ratio is inverted to determine a 2$\sigma$ upper limit for each signal mass.
The excluded region from this analysis is shown in figure \ref{fig:hps15limit}.
This result only used about 1\% of the approved run time of the experiment. 

\begin{figure}[]
\centering
    \includegraphics[height=2.2in]{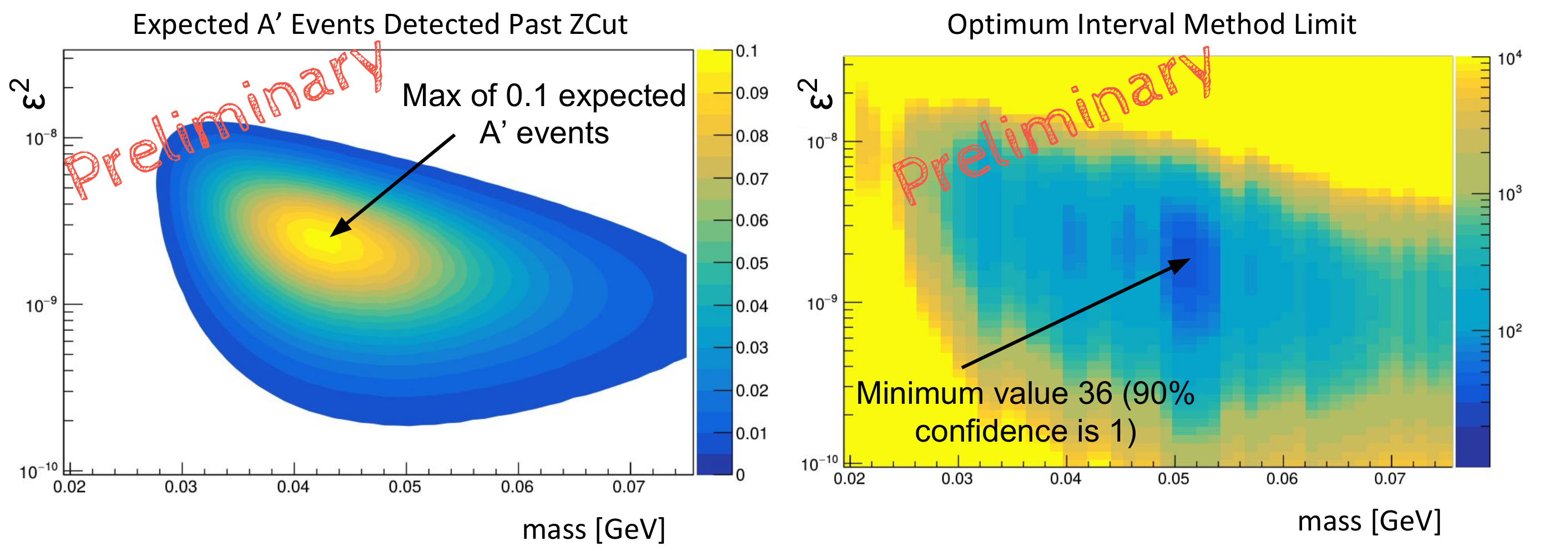}
    \caption{The signal expectation and limit set using the optimum interval method for the 2015 dataset.
    It can be seen that at the maximum only about 0.1 signal events are expected.}
\label{fig:hpsVtxResult}
\end{figure}

A displaced vertex search is currently being performed on the 2016 dataset. Preliminary results
of this search analysis were presented at the APS April Meeting 2019 \cite{APS19}. This search
reconstructs the position of the $e^+e^-$ vertices and searches for an excess of vertices
downstream of the target. The total signal expected for the maximum signal point is on the order
of a tenth of an event, so there is not enough luminosity to set any limits using the 2015 dataset.
The preliminary result of this analysis is shown in figure \ref{fig:hpsVtxResult}.

\subsection{HPS Upgrades and 2019 Physics Run}
\begin{figure}[]
\centering
    \includegraphics[height=2.3in]{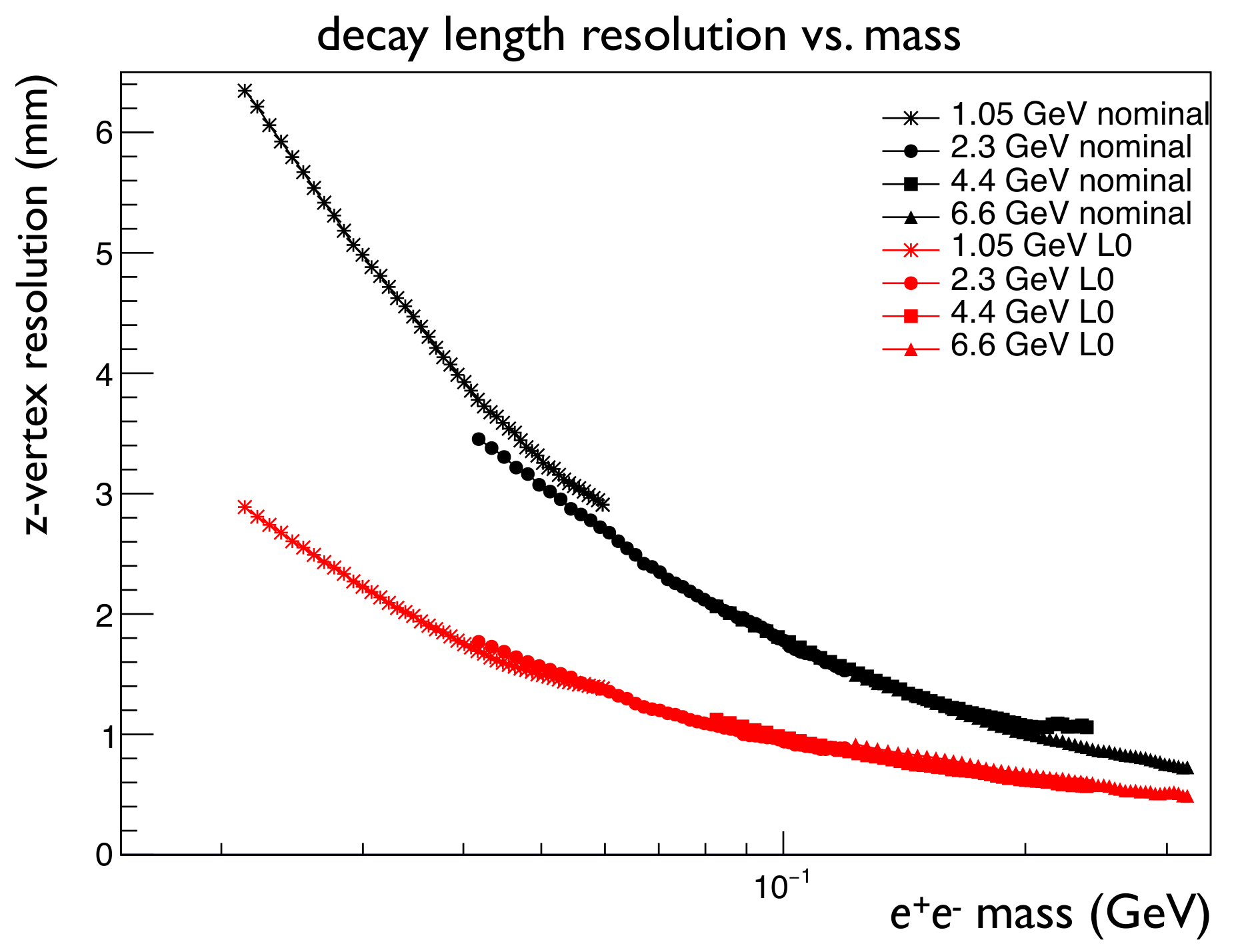}
    \caption{The decay length resolution in MC with and without the L0 upgrade.}
\label{fig:svtResImp}
\end{figure}

\begin{figure}[]
\centering
    \includegraphics[height=2.3in]{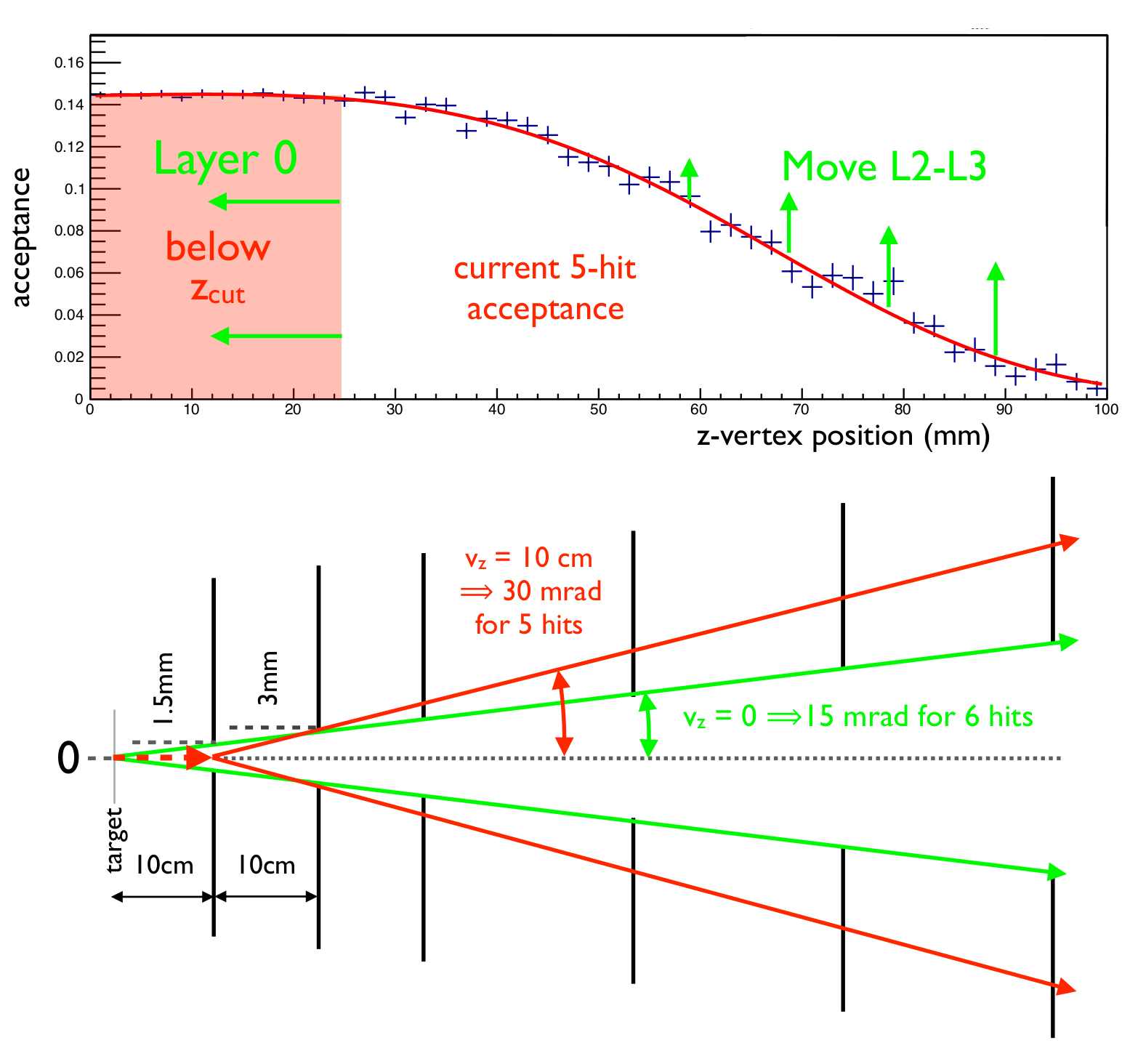}
    \caption{The acceptance of signal as a function of vertex position, showing how the
    addition of L0 and repositioning of Layers 2 and 3 will increase the acceptance. Below
    we see a cartoon depiction of how the new geometry improves this performance.}
\label{fig:SVTupMot}
\end{figure}

\begin{figure}[]
\centering
    \includegraphics[height=5.5in]{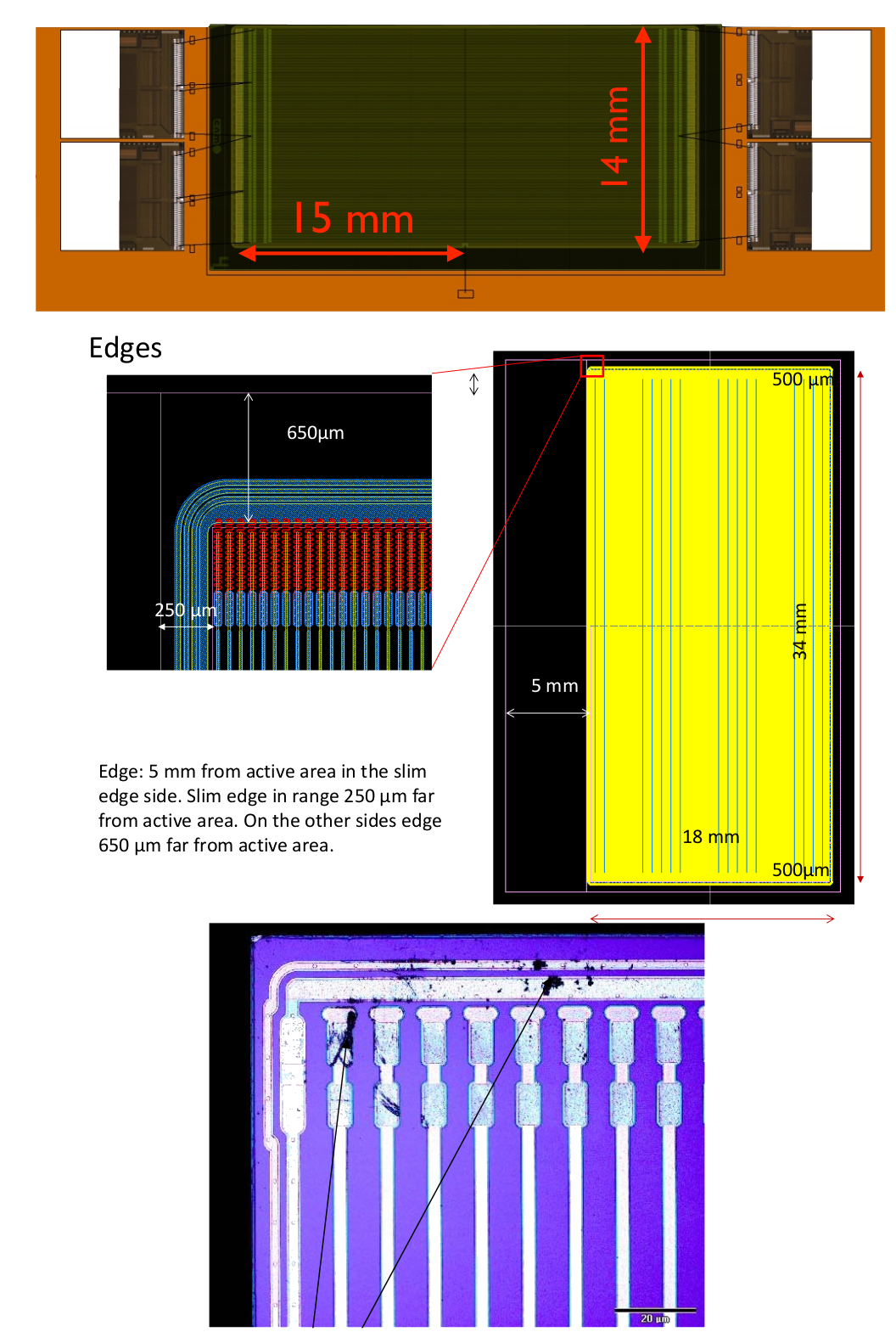}
    \caption{A collection of drawing and pictures of the new L0 sensor technology.}
\label{fig:L0sensors}
\end{figure}

Two significant upgrades were executed on the HPS apparatus before the 2019 physics run.
The first is to the SVT, adding a new "Layer 0" (L0) which is closer to the target, allowing
access to shorter decay lengths which significantly increases the acceptance of the experiment.
The improvement in vertex resolution with the addition of L0 is shown in figure \ref{fig:svtResImp}.
Also, the original first layer was replaced with the same new technology used in L0, which
moves the active region of the detector closer to the beam. Layers 2 and 3 were also moved closer 
to the beam as occupancy allowed, ultimately increasing the acceptance at longer decay lengths.
Figure \ref{fig:SVTupMot} shows how the upgrades are expected to improve the performance of the experiment.
The technology of the new L0 sensors has about half the material, being 200 microns thick.
It has a 55 micron readout pitch split into two 15 mm by 14 mm active areas, to decrease the per-strip
occupancy. The inactive region near the beam is only 250 microns, allowing closer placement of
the active region to the beam. Replacing Layer 1 with this new technology
will also decrease difficult backgrounds such as wide-angle bremsstrahlung photons converting 
in the inactive region of the sensor and trident production in this region from scattered electrons. 
This new sensor technology is produced by CNM Barcelona and is shown in figure \ref{fig:L0sensors}.

\begin{figure}[]
\centering
    \includegraphics[height=2.8in]{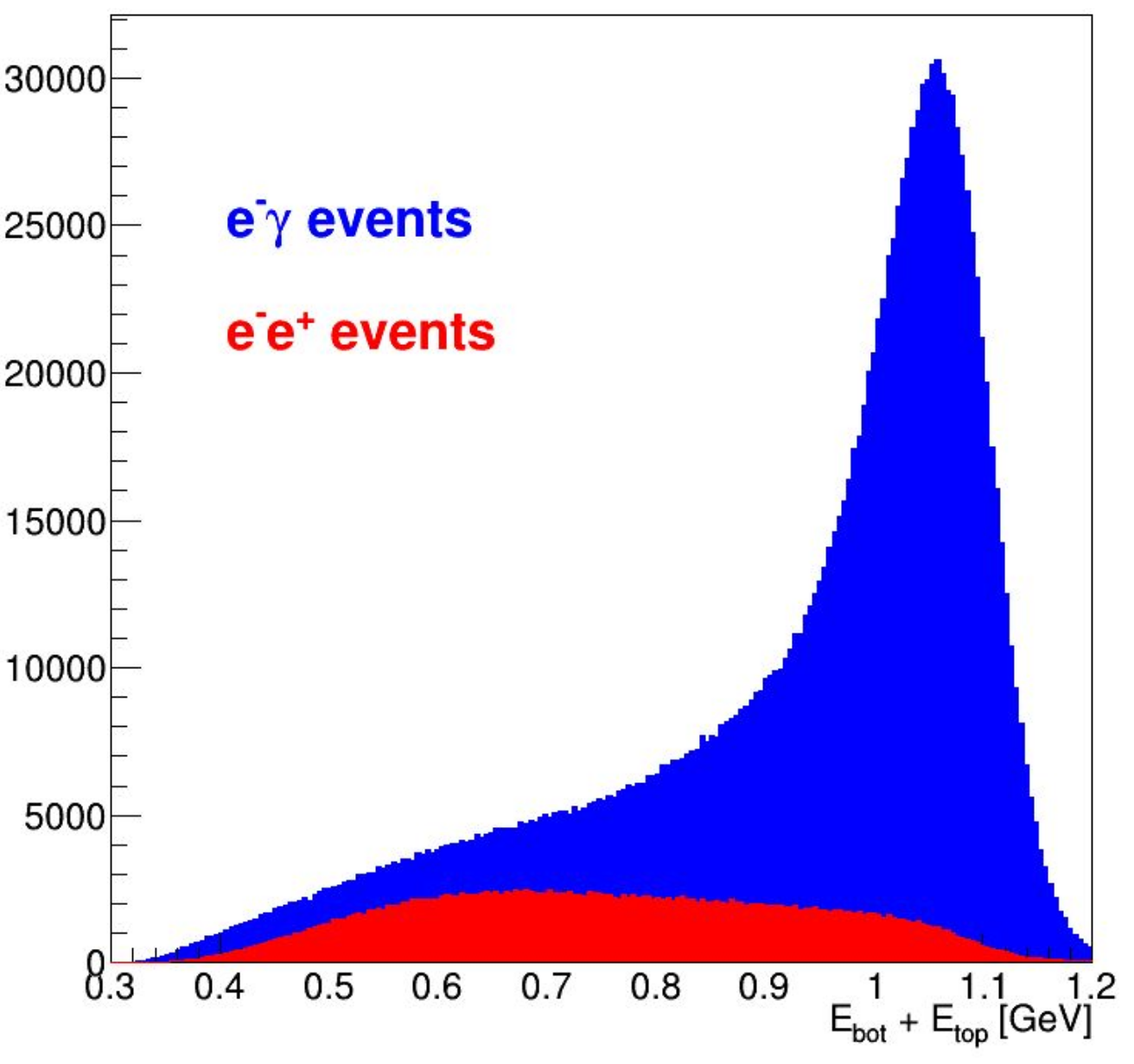}
    \caption{The number of events triggered events with photons and
    events with positrons. The blue region is suppressed in the trigger with the hodoscope
    upgrade.}
\label{fig:trigGamma}
\end{figure}

The other upgrade to the apparatus is the addition of a hodoscope for triggering on positrons.
This allows for a single cluster trigger to be used for physics production data taking, since
the hodoscope will allow for the reduction of photons as shown in figure \ref{fig:trigGamma}.
The hodoscope uses two layers of scintillating tiles with a wavelength-shifting fiber that
carries the light out to a multichannel PMT. The trigger requires two layers of hodoscope
hits with a position that matches the position and time of the cluster in the ECal.

\begin{figure}[]
\centering
    \includegraphics[height=2.5in]{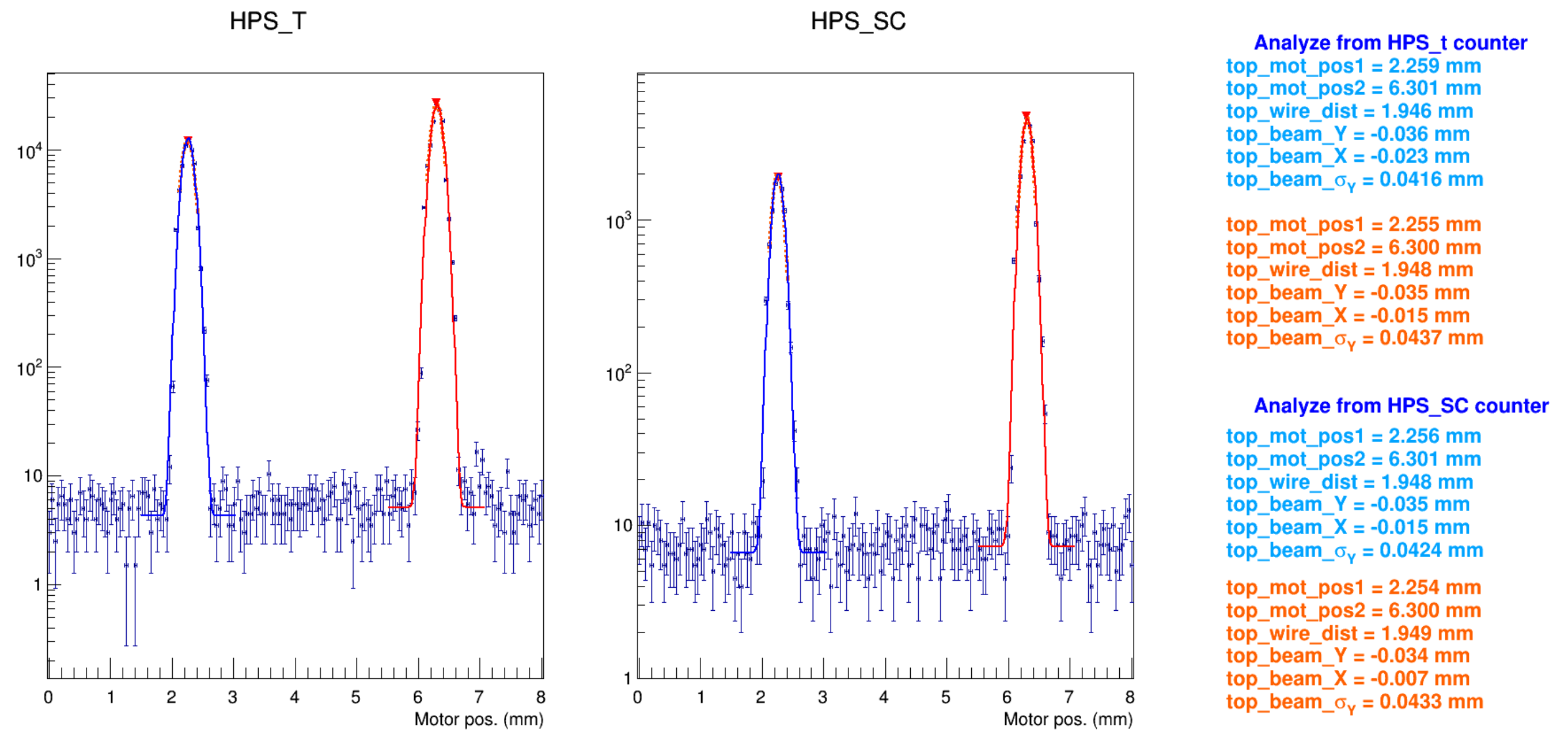}
    \caption{An example of an analyzed SVT wire scan. Here we see a beam spot size of about 42 um
    within 50 um of the center of the detector. This demonstrates the CEBAF is capable of delivering
    physics production quality beam to the experiment.}
\label{fig:hpsBeamPro}
\end{figure}

\begin{figure}[]
\centering
    \includegraphics[height=2.8in]{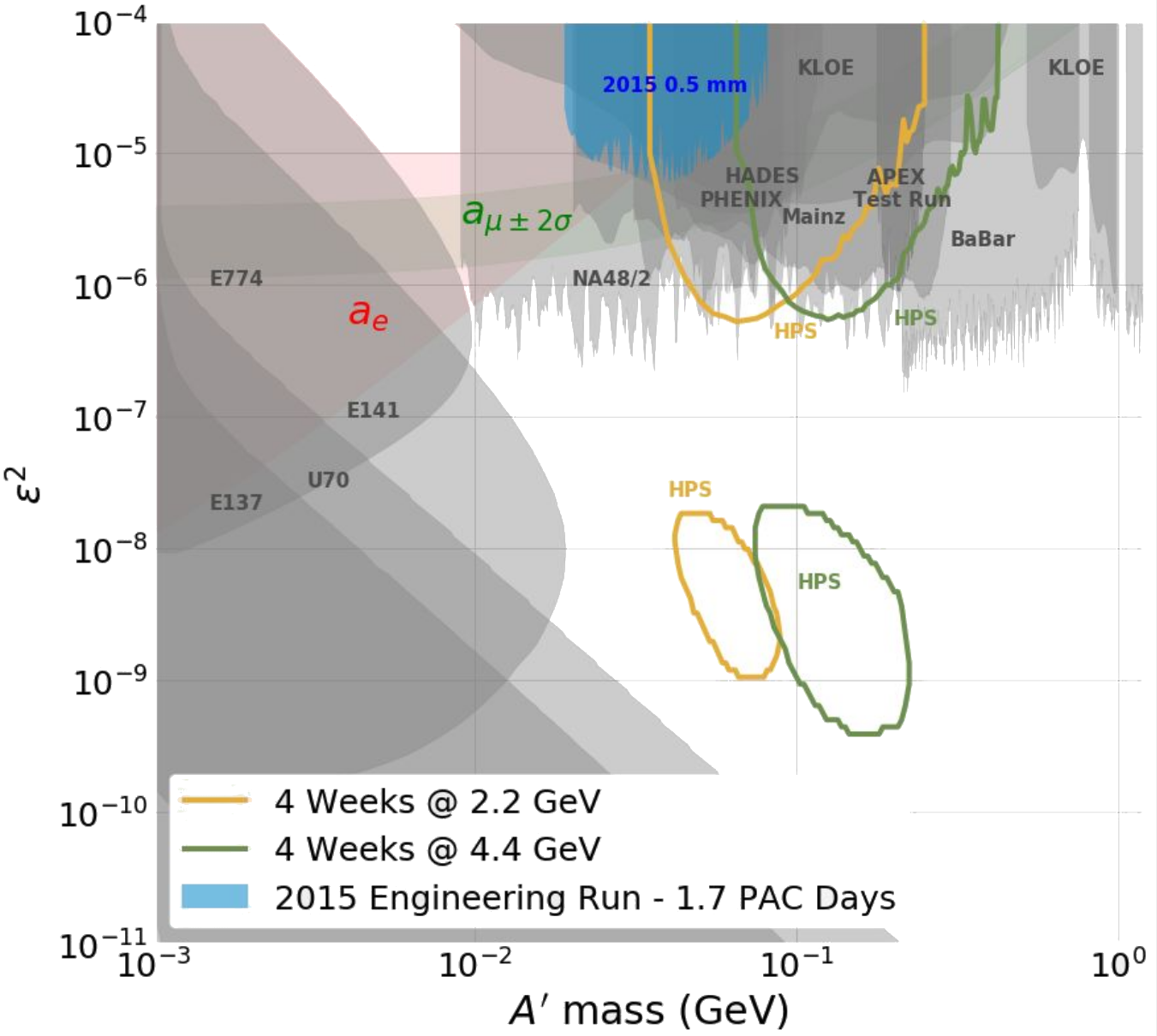}
    \caption{The reach estimate for the run in the summer of 2019 for two different beam energies.
    The 2.2 GeV estimate assumes a 200 nA beam incident on a 4 um W target. The 4.4 GeV estimate assumes
    a 300 nA beam incident on an 8 um W target. The 2019 data run is at a beam energy of about 4.5 GeV.}
\label{fig:reach19}
\end{figure}

HPS is a challenging experiment for the CEBAF to deliver the required beam to. The SVT has wires
installed that allow the determination of the beam position and size at the target location.
An example of one such scan is shown in figure \ref{fig:hpsBeamPro}. The CEBAF has demonstrated
the ability to deliver beam within the acceptable specifications for physics data production.
Squeezing the beam to an adequate size on the target can take many hours of tuning work.
After the completion of the upgrade program, the HPS reach has been re-evaluated. The new
reach estimates for the 2019 data run are shown in figure \ref{fig:reach19}.

\section{Conclusion}
The HPS detector has performed well for the 2015 and 2016 engineering runs, producing results that establish the physics performance of the experiment. The preliminary vertex analysis demonstrates that new sensitivity is expected with larger datasets, beginning with data from the Summer 2019 run. HPS underwent a successful upgrade program before this run to improve the reach of the experiment.  Because of the extreme proximity of the silicon to the primary beam, the HPS experiment presents major challenges to beam delivery by CEBAF, requiring extensive beam tuning during operations. The 2019 run is currently on-going and should produce a dataset which allowing for sensitivity to A$^\prime$ with mm-to-cm decay lengths for the first time.

\section*{Acknowledgements}
Thanks to the entire HPS collaboration for their support in making this presentation.
Special thanks to Tim Nelson, Omar Moreno, Matt Solt, Matthew Graham, and 
Rafayel Paremuzyan for producing most of the figures used in this presentation.


\begin{thebibliography}{99}


\bibitem{WIMP}
Lee B W and Weinberg S 1977 Phys. Rev. Lett. 39 165

\bibitem{Holdom}
Bob Holdom, Physics Letters B, Volume 166, Issue 2, 1986, Pages 196-198

\bibitem{ECalNIM}
I. Balossino et al. NIM A854, p. 89-99 (2017)

\bibitem{aprime}
R. Essig, P. Schuster, N. Toro et al. J. High Energ. Phys. (2011) 2011: 1. https://doi.org/10.1007/JHEP02(2011)009

\bibitem{apv25}
L.L. Jones et al., The APV25 deep submicron readout chipfor CMS detector, Proceedings of the Fifth Workshop onElectronics for LHC Experiments, CERN 99-09, CERN/LHCC/99-33, pp. 162–166, http://hep.physics.wisc.edu/LEB99.

\bibitem{CosmicVisions}
Cosmic Visions Whitepaper, arXiv:1707.04591

\bibitem{res15}
P. H. Adrian et al. (Heavy Photon Search Collaboration), Phys. Rev. D 98, 091101(R)

\bibitem{APS19}
M. Solt, The Heavy Photon Search Experiment, APS April Meeting 2019, http://meetings.aps.org/Meeting/APR19/Session/J09.1

\end{thebibliography}
\end{document}